\documentclass[aip,jmp,preprint,amsmath,amssymb,showpacs,superscriptaddress,12pt]{revtex4-1} 
\usepackage{amsmath,amssymb,graphicx}
\usepackage{graphicx}
\usepackage{dcolumn}
\usepackage{bm}
\begin{document}

\title{\Large{A protocol based on compressed sensing for high-speed authentication and cryptographic key distribution over a multiparty optical network}}

\author{Wen-Kai Yu}
\affiliation{\small{Laboratory of Space Science Experiment Technology, Center for Space Science and Applied Research, Chinese Academy of Sciences, Beijing 100190, China}}
\affiliation{\small{University of Chinese Academy of Sciences, Beijing 100049, China}}

\author{Shen Li}%
\affiliation{\small{Laboratory of Space Science Experiment Technology, Center for Space Science and Applied Research, Chinese Academy of Sciences, Beijing 100190, China}}%
\affiliation{\small{University of Chinese Academy of Sciences, Beijing 100049, China}}

\author{Xu-Ri Yao}
\affiliation{\small{Laboratory of Space Science Experiment Technology, Center for Space Science and Applied Research, Chinese Academy of Sciences, Beijing 100190, China}}%
\affiliation{\small{University of Chinese Academy of Sciences, Beijing 100049, China}}

\author{Xue-Feng Liu}
\affiliation{\small{Laboratory of Space Science Experiment Technology, Center for Space Science and Applied Research, Chinese Academy of Sciences, Beijing 100190, China}}%
\affiliation{\small{University of Chinese Academy of Sciences, Beijing 100049, China}}

\author{Ling-An Wu}\email{wula@aphy.iphy.ac.cn}
\affiliation{\small{Laboratory of Optical Physics, Institute of Physics and Beijing National Laboratory for Condensed Matter Physics, Chinese Academy of Sciences, Beijing 100190, China}}%

\author{Guang-Jie Zhai}
\affiliation{\small{Laboratory of Space Science Experiment Technology, Center for Space Science and Applied Research, Chinese Academy of Sciences, Beijing 100190, China}}%

\begin{abstract}
We present a protocol for the amplification and distribution of a one-time-pad cryptographic key over a point-to-multipoint optical network based on computational ghost imaging and compressed sensing (CS). It is shown experimentally that CS imaging can perform faster authentication and increase the key generation rate by an order of magnitude compared with the scheme using computational ghost imaging alone. The protocol is applicable for any number of legitimate users, thus the scheme could be used in real intercity networks where high speed and high security are crucial.
\end{abstract}


\maketitle 

\section{Introduction}
In cryptographic communication systems, conventional key distribution protocols either rely on trusted couriers or are based on the principles of public-key cryptography. The former is unpractical and costly for large systems, while the latter relies on the difficulty of solving certain mathematical problems within a reasonable time. For the provenly secure one-time-pad the key must be perfectly random and as long as the message, which makes it very cumbersome.

Quantum key distribution (QKD), first proposed by Bennett and Brassard in 1984, is guaranteed by Heisenberg's uncertainty principle to provide absolute security \cite{Bennett1984, Bennett1992, Brassard1992, Ekert1992}, and has the unique advantages in that it can detect eavesdropping \cite{Brassard1992}. However, the key generation rate and transmission distances are limited due to photon losses, so QKD cannot be deployed over arbitrarily long distances without the help of quantum repeaters, which are extremely costly and not feasible at present. Even if no photons were lost, due to the inherent nondeterministic outcome of the projection of quantum states, 50\% of the bits would have to be discarded and wasted in the well-known BB84 protocol \cite{Bennett1984}. To overcome this, certain variations of the latter have been proposed, such as Lo's scheme to use asymmetric probabilities for the transmission and measurement bases combined with separate error analysis of the two subsets, which in the asymptotic limit could double the key generation rate to 100\% \cite{Lo2005}. In another variant proposed by Hwang et al. \cite{Hwang1998}, Alice and Bob use a common secure start random sequence to set their bases, which consequently do not have to be publicly compared and so no qubits need to be discarded; although some previously shared secure information is required, it could be used repeatedly so long as the generated key has not been used to encode and send a message. Another scheme called quantum key evolution by Guan et al. \cite{Guan2012} also requires Alice and Bob to first establish a common start key by the BB84 protocol to encode a message, from which the key is updated, i.e. a new key is generated, by means of error correction and hashing every time a message is sent through a quantum channel, the advantage being that overall fewer qubits are required for sending long messages. However, all the above schemes only improve the efficiency in traditional QKD or long message transmission, and still require a quantum channel to transmit qubits so they are just as susceptible to the frailty of photon losses. There is no direct amplification of the key, and the longstanding problem of multiparty key distribution is also not addressed. Classically, a true random number sequence could be used as a seed to generate a large key with sufficient randomness by means of some computer program, but again secure multiparty distribution of this new long key is not so trivial.

Ghost imaging (GI) is an intriguing imaging technique that exploits the nature of light correlations. The first experimental demonstration of ghost imaging was performed using entangled photons generated by spontaneous parametric downconversion \cite{Pittman1995}. Soon after, it was shown that GI is achievable also with pseudothermal \cite{Lugiato2004} or true thermal light \cite{LAWu2005}. A theory of GI was developed in a Gaussian-state framework that provided a general formalism for both classical and quantum GI \cite{Erkmen2008}, while Brida et al. presented a systematic analysis of the signal-to-noise ratio in such systems \cite{Brida2011}. In pseudothermal GI light from a laser source passes through a random diffuser, usually a rotating ground-glass plate, and then is split by a beamsplitter into an object and a reference arm. In the former, a bucket (single-pixel) detector is placed behind the object, yielding no spatial information but merely measuring the total light intensity impinging on it. However, in the reference arm there is a detector with spatial resolution, which records images of the random light field coming from the rotating diffuser. In order to reconstruct an image of the object, the outputs of the two detectors must be correlated. Recently, Shapiro \cite{Shapiro2008} proposed a modified version of GI called computational ghost imaging, where the reference arm is dispensed with and random speckle fields are generated by a spatial light modulator. This scheme has already been used for optical encryption of pictorial objects \cite{Clemente2010, Tanha2012}, but the security is questionable as a blurred image may still be reconstructed if an attacker knows a certain percentage of the frames.

The development of compressed sensing (CS), first proposed by Donoho \cite{Donoho2006} and Cand$\grave{e}$s (who called it compressive sampling) \cite{Candes2006} in 2006, gave birth to a new era of signal processing and information theory. Their work shows that signals can be acquired from non-adaptive measurements with a sampling rate much lower than the Nyquist-Shannon limit \cite{Donoho2006, Candes2006}. An important application is the single-pixel camera proposed and demonstrated by Baraniuk's group \cite{Baraniuk2008, Daniel2008} in a THz imaging setup, which was followed soon after by the compressive GI experiment of Silberberg and coworkers \cite{Silberberg2009}.

\begin{figure}
\includegraphics[width=8.5cm]{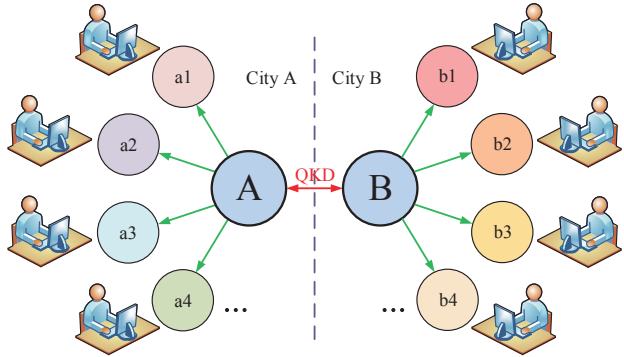}
\caption{\label{fig:network} (Color online) Secure communication between two cities.}
\end{figure}

In a previous work \cite{LiShen2013} we demonstrated a protocol for privacy amplification and point-to-multipoint key distribution in an optical network based on computational GI. The method is purely classical, but may be used to amplify a random start key generated originally by any one of the standard quantum key distribution protocols. The stochastic intensity fluctuation of the thermal light ensured the true randomness of the key generated, which used a secure measurement matrix as a start key. In this paper we use CS imaging for faster authentication, while the key generation rate can be increased by an order of magnitude. A possible network system is illustrated in Fig.~\ref{fig:network}, where keys can be distributed between City A and City B through a QKD channel, then shared with any number of users in each city network by our point-to-multipoint key distribution protocol.

\section{Compressed sensing}
In CS, a natural signal $x \in {\Re^{N}}$ can be sparsely represented in a certain basis $\Psi$ (e.g. Haar wavelet basis). Mathematically, this can be described by:
\begin{eqnarray}
y = Ax + e,   x = \Psi x',
\end{eqnarray}
where $y \in {\Re^{M}}$ is a measured vector, $A \in {\Re^{M \times N}}$ is a measurement matrix, and $e$ is the noise. Rows of $A$ should be incoherent with the sparse basis. It is found that a completely random $A$ is best, from which accurate recovery is possible by convex optimization \cite{Candes2006}:
\begin{eqnarray}
\mathop {\min }\limits_{x'} \frac{1}{2}\left\| {y - A\Psi x'} \right\|_2^2 + \tau {\left\| {x'} \right\|_1},
\end{eqnarray}
where $\tau$ is a constant scalar, and $\left\|  \cdot  \right\|_p$ stands for $l_p$ norm.

\section{Ghost imaging}
As a matter of fact, the mathematical model for computational GI can also be described by Eq.~(1), where $A$ consists of $M$ random frames, and the intensity distribution of each frame on the spatial light modulator is ${a_i}\left( {m,n} \right)$; $m, n$ are the row and column lengths of ${a_i}$, with coordinates $c$ and $d$, respectively. The vector $y$ consists of many $y_i$, which correspond to the total intensities measured by the bucket (single-pixel) detector in the $i$th exposure. Once we know the matrix $A$ and all the intensities in $y$, the final image can be reconstructed by GI. In this work we shall use differential ghost imaging (DGI) \cite{Gatti2010} to recover the image:
\begin{eqnarray}
G_{\textrm{DGI}}\left({m,n}\right)=&\left\langle{{y_i}{a_i}\left({m,n}\right)}\right\rangle-\frac{{\left\langle{{y_i}}\right\rangle}}{{\left\langle B\right\rangle}}\left\langle{{a_i}\left({m,n} \right)B}\right\rangle,
\end{eqnarray}
where
\begin{eqnarray}
B=\sum\limits_{c = 1}^m {\sum\limits_{d = 1}^n {{a_i}\left( {c,d} \right)} }.
\end{eqnarray}

\section{Our improved protocol}
We note that the gray value of each pixel of the reconstructed image will be defined once $A$ and $y$ are fixed. To obtain a random number series for the cryptographic key, we extract certain digits after the decimal point of each gray value; the random quality of the series will depend on the randomness of both $A$ and the light source.

The basic steps in our improved protocol are as follows.\\
1.	The server distributes start keys which correspond to a measurement matrix $A$, and other essential parameters to the legitimate parties through a QKD channel or some other secure approach (e.g. flash cards) in advance. The essential parameters include which object masks are to be used, the imaging area, and the number of measurement frames.\\
2.	The legitimate parties tell the server their user names at the beginning of their multiparty communication; this can be over a public channel.\\
3.	The server measures the bucket (single-pixel) detector intensities behind the object mask, and sends these values via an open channel to all the legitimate parties simultaneously, according to their user names.\\
4.	The parties retrieve the image via a CS algorithm to check whether the data has been falsified by any potential eavesdropper and for server authentication as well. If the reconstructed image matches the preselected object, then the parties and server may safely conclude that there has been no tampering of the public messages.\\
5.	The parties then reconstruct their images from the $G_{\textrm{DGI}}\left({m,n}\right)$ of Eq. (3).\\
6.	Certain digits after the decimal point of each gray value are selected to form a binary sequence, 0 for an even number, and 1 for an odd number. These digits do not contribute visibly to the gray tone of the image.\\
7.	The parties and the server test for eavesdropping by publicly comparing bits of a random subset of the key on which they should all agree. If there are no discrepancies or the bit error rate does not exceed $\sim$0.08\%, they can conclude that the remaining uncompared data are secure enough for use as a one-time-pad key, otherwise it may be suspected that Eve has interfered, in which case the server and legitimate parties should discard the generated keys and return to Step 2.

In an actual system, channel noise will also inevitably give rise to bit errors. It is difficult to differentiate whether the errors are caused by the noise or illegal intrusion, thus it is necessary to ensure that all parties have the same secret key by keeping the discrepancies under a tolerable level.

Next time, the server only needs to measure new values of $y$ for different preselected objects, without having to change the start matrix $A$. Although the randomness is guaranteed by the thermal light, the security might be compromised when the same start key matrices are repeatedly used too many times, eventually leading to some periodicity to occur. In practice, the original start keys as well as the object masks should be replaced from time to time, but in principle it could be said that a one-time distribution of $A$ can generate as many one-time pad keys as required, independent of the key distribution rate and transmission distance, which is a great advantage in very high security applications such as intercity secure communication, efficient secret conferencing, and electronic payment.

\section{Experimental results and analysis}
\begin{figure}
\includegraphics[width=8.5cm]{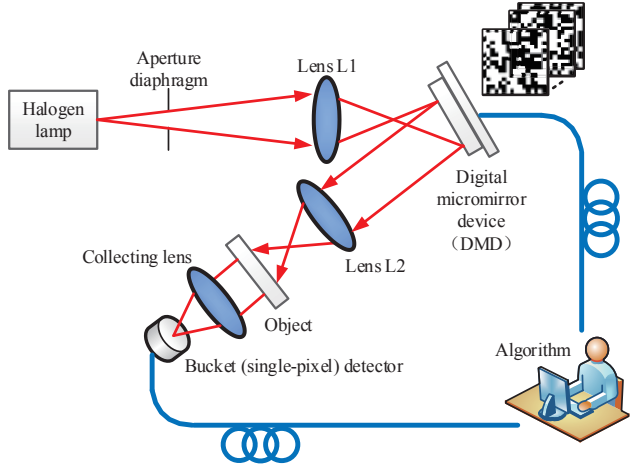}
\caption{\label{fig:apparatus}(Color online) Experimental setup.}
\end{figure}

A schematic of the experimental setup is shown in Fig.~\ref{fig:apparatus}. Unfiltered thermal light from a 55~W halogen lamp passes through an aperture diaphragm and lens L1 onto a digital micromirror device (DMD), which functions as a spatial light modulator but only produces two intensity modulation states. The DMD array consists of $1,024 \times 768$ micromirrors of size $13.68 \times 13.68$~$\mu$m. Each mirror rotates about a hinge and can be positioned in one of two states, + or $-12^\circ$ from its initial position, as determined by the random binary program loaded into the controller. Each DMD pattern corresponds to an array ${a_i}\left( {m,n} \right)$. The light reflected from the mirrors at $+12^\circ$ is focussed by a second lens L2 onto the object on a translation stage, then transmitted and focussed by a collecting lens onto a bucket (single-pixel) detector, which records the total light intensity ${y_i}$. Using these values of ${y_i}$ sent by the server, the receiving parties then use the TVAL3 algorithm \cite{Li2010} for identity authentication, and DGI reconstruction for key generation.

\begin{figure}
\includegraphics[width=9.5cm]{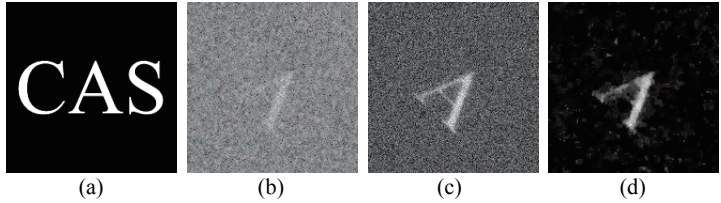}
\caption{\label{fig:M} (a) Image of the original mask. (b), (c), and (d): Images of the letter ``A" reconstructed experimentally by: DGI, with $M=1000$; DGI, with $M=13500$; CS, using the same experimental data as in (b), MSE$=4.642 \times 10^2$.}
\end{figure}

In the experiment, a black-and-white film with transparent letters ``CAS" is used as the object, and Fig.~\ref{fig:M} (a) shows its original photograph taken by a 1/1.8 in. charge-coupled device (CCD) with $1,280 \times 1,024$ pixels and an exposure rate of 26 frames/s. A start key is then used as a measurement matrix to control the DMD, which is set to a working area of $160 \times 160$ micromirrors (pixels). Here the same CCD camera will be used as the bucket (single-pixel) detector, by integrating the gray values of all the pixels in each exposure. After that, these measured total values are transmitted to the parties concerned through the public channel.

Knowing the start key, secure parameters and transmitted measured values, it is simple for each party to obtain the required secret key by DGI. Usually, the calculated gray values of the recovered image are accurate within a certain range. Some least significant digits after the decimal point are converted into a binary sequence to generate keys according to their parity. Actually, these digits after the decimal point just represent tiny random fluctuations of the noise rather than useful object information. It is known that random numbers can be created by various physical processes, and here the measured values reflect total light intensity fluctuations which include the light source fluctuations, scattering and absorption noise along the optical path, and electrical noise. Assuming that the measurement start matrix was also generated by a random physical process, the key obtained should be completely random. To check this, we use the common ENT program \cite{Walker2008}. In the test results, if the entropy is 1 bit per bit, arithmetic mean value 0.5, Monte Carlo value for $\pi$ close to $\pi$, and the serial correlation coefficient is 0, we can conclude that the binary sequence is completely random.

Next, we restricted the object region to just the letter ``A" but with the same DMD area, and changed the number of exposures for each image, varying from 1,000 to 13,500. Each group of measurements was repeated 5 times, to produce binary sequences of length 128,000 bits. The DGI image reconstructions are shown in Figs.~\ref{fig:M} (b) and (c), and the results of their randomness tests in Table~\ref{tab:table2}. We can see that the number of measurements has little effect on the randomness of the generated keys.

In order to evaluate quantitatively the quality of the images $G_{\textrm{DGI}}\left({m,n}\right)$, we calculate the mean square error (MSE) of the reconstructions compared to the original object $X_{\textrm{ori}}$:
\begin{eqnarray}
\textrm{MSE} = \frac{1}{{mn}} \cdot \sum\limits_{c=1}^m {\sum\limits_{d = 1}^n {{{\left( {{X_{\textrm{ori}}}\left( {c,d} \right) - { G_{\textrm{DGI}}}\left( {c,d} \right)} \right)}^2}} }.
\end{eqnarray}
Naturally, the smaller the MSE, the better the quality of the image.
\begin{table}
\caption{\label{tab:table2}ENT randomness test results for different numbers of frames.}
\begin{center}
\begin{tabular}{ccccc}\hline\hline
    $M$ & Entropy & Mean & $\pi$ & Correlation\\ \hline
    1,000 & 1.000000 & 0.4998 & 3.167292 & 0.001062\\\hline
    3,500 & 1.000000 & 0.4995 & 3.126782 & 0.004749\\\hline
    6,000 & 1.000000 & 0.4998 & 3.137284 & 0.002625\\\hline
    8,500 & 1.000000 & 0.4996 & 3.179295 & 0.001624\\\hline
    11,000 & 1.000000 & 0.5001 & 3.129782 & 0.002344\\\hline
    13,500 & 0.999992 & 0.5017 & 3.122280 & -0.002698\\\hline\hline
\end{tabular}
\end{center}
\end{table}

In Fig.~\ref{fig:M} (b), we can hardly distinguish the letter ``A" and the $\textrm{MSE}=4.953 \times 10^6$ is also high, for the number of exposures is only 1,000. In Fig.~\ref{fig:M} (c), 13,500 exposures were used to reconstruct the image, which is much improved with the $\textrm{MSE}=4.192 \times 10^5$. Although the image quality can be improved by increasing the number of exposures $M$, the latter has little effect on the randomness of the key. In Fig.~\ref{fig:M} (d) the image was recovered by CS and is visibly much clearer than in (c) although only 1,000 exposures were used, the same as in (b). Generally, when the MSE value exceeds $10^3$ for the figures that are automatically gray-scale compensated, the legitimate parties can tell whether the CS image is different from the preselected object, in which case they can judge that the keys generated are not secure enough. This demonstrates the great advantage of CS which is particularly effective in the case of under-sampling. Actually, DGI and CS are fundamentally different, as the former can be regarded as an algorithm for normalized weighted averaging, while the latter can be seen as an algorithm for convex optimization. A sequence rearrangement of ${a_i}\left( {m,n} \right)$ and $y_i$ pairs will not affect the DGI image but will severely influence the results of CS. If a legitimate party finds that the image reconstructed by CS is different from the original object, he/she can immediately suspect tampering by an eavesdropper, thus CS reconstruction can play a valuable role in identity authentication. The great advantage of CS is in removing noise and improving image quality, so it can retrieve better images than DGI with fewer exposure frames. Thus for the black-and-white masks used in our proof-of-principle experiment we can see that the contrast is much sharper and the background of the images recovered by CS is much blacker; that is, the background values are almost all zero, which would give a much higher ratio of 0’s in the bits extracted from that part of the image. This, however, results in worse randomness than from DGI imaging, which is why we do not use CS but DGI imaging in generating the keys (the quality of the DGI image is unimportant here), while CS imaging is used only for high-speed authentication. In real applications, instead of binary masks, complex objects with rich grey tones should be used, which in traditional ghost imaging would require an enormous number of frames to reconstruct but with CS can be quickly recovered, increasing the DGI key generation rate by an order of magnitude. For $M=1,000$, the total length of the start key consumed was $1,000 \times 160 \times 160=25,600,000$, and a one-time-pad key of length $160 \times 160=25,600$ was generated each time, but the process could be repeated indefinitely many thousands of times, to realize privacy amplification. As the number of exposures has little effect on the randomness of the key, the fewer the exposures used to recover the image for authentication, the higher the key generation rate. However, decreasing the number of frames will increase the probability for the image to be deciphered. Actually, for one frame of size $160 \times 160$, it will take a top supercomputer of speed 33.86 Pflops $10^{7,679,976}$ years to guess 1,000 frames. In our previous work \cite{LiShen2013} where conventional GI was used for image reconstruction, a blurred image could be seen only when the number of frames exceeded 20,298. This can also be used for authentication but is rather slow, and the key generation rate is low. With our CS algorithm, 1,000 frames are sufficient for authentication, thus the key generation rate is increased by an order of magnitude. With current commercial technology \cite{TI1080, TI4100}, the size of the DMD could be $1,920 \times 1,080$ pixels and the frame rate up to 23 kHz, so the bit rate could reach several Gbit/s. Furthermore, high-order correlations could be used for acquisition of sharper image quality compared with conventional second-order GI schemes.

\begin{figure}
\includegraphics[width=12.5cm]{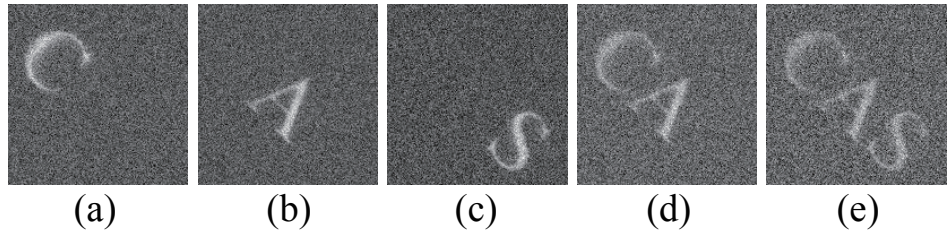}
\caption{\label{fig:region} Images reconstructed experimentally by DGI for different regions. (a), (b), (c), (d) and (e): reconstructed DGI images of the letters ``C", ``A", ``S", ``CA", and ``CAS", respectively.}
\end{figure}

To analyze the influence of different object areas and images on the randomness of the key, we block part of the mask, allowing light to pass through those regions containing the letters ``C", ``A", ``S", ``CA", and ``CAS". However, the images reconstructed from the total gray value of each region are derived from the same $160 \times 160$ size array on the DMD, so the generated secure key is still 25,600 bits in length although the number of frames used for reconstruction was only 10,149. The DGI reconstructed images are shown in Figs.~\ref{fig:region} (a)-(e). All experiments were repeated 10 times, and the data analyzed by the ENT tests. Table~\ref{tab:table1} shows that the extracted digits are random enough for a one-time-pad key, and that the randomness is independent of the region selected, thus any object mask can be regarded as a secure parameter.
\begin{table}
\caption{\label{tab:table1}ENT randomness test results for different letter images.}
\begin{center}
\begin{tabular}{ccccc}\hline\hline
    Region & Entropy & Mean & $\pi$ & Correlation\\ \hline
    ``C" & 0.999999 & 0.5005 & 3.129196 & 0.000015\\\hline
    ``A" & 0.999999 & 0.5006 & 3.166698 & 0.001280\\\hline
    ``S" & 0.999995 & 0.5014 & 3.131446 & -0.000008\\\hline
    ``CA" & 1.000000 & 0.4998 & 3.143446 & -0.003203\\\hline
    ``CAS" & 0.999985 & 0.4978 & 3.162948 & -0.003270\\\hline\hline
\end{tabular}
\end{center}
\end{table}

\section{Security analysis}
We now examine the security of our multiparty key distribution scheme against various forms of attack. The essence of the protocol is that a random one-time-pad is amplified by a truly random function, i.e. the purely stochastic intensity fluctuations of the thermal light source. Let us now consider how an eavesdropper Eve could attempt to break the code, assuming that under conditions which do not violate the basic laws of physics, she possesses any advanced capability and means.

1) Eavesdropping

In the public channel, Eve has no problem intercepting the measurement values. Even if she knows all the measurement values, to recover the key she must guess all the frames and secure parameters. Since they are distributed via a QKD system or some other totally safe method in advance, it is impossible for her to intercept or gain any knowledge of the start key. For the start key, assuming that its size is $1,000 \times 25,600$ where 1,000 is the number of frames and 25,600 is the length of each frame of size $160 \times 160$, and that its binary matrix elements exhibit a random distribution, the probability of its being deciphered by Eve through exhaustive attack is $2^{-25,600,000}$.

Again, assuming that we generate $160 \times 160$ bits in each communication and use just 1,000 of them as a one-time-pad key, Eve must sweep through $A_{25,600}^{1,000}$ possible permutations. A simple calculation reveals that it will take too many years to finish this search. Moreover, since each run is totally different, her efforts will turn out to be a futile attempt.

\begin{figure}
\includegraphics[width=14.5cm]{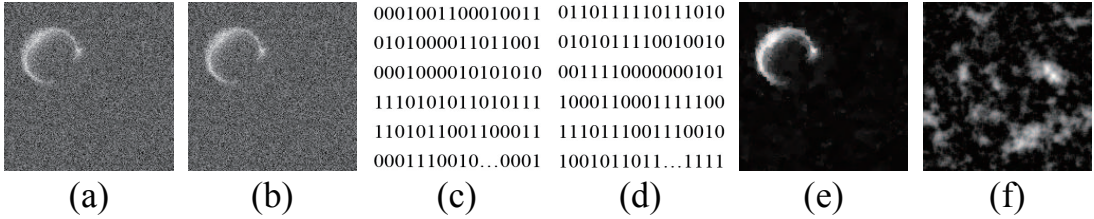}
\caption{\label{fig:comparison} (a) DGI image recovered from the correct start key; (b) DGI image recovered with one bit of the start key incorrect; (c) and (d), the keys obtained from (a) and (b), respectively. (e) CS image retrieved from the correct start key; (f) CS image retrieved with 19 bits of the measured intensity incorrect.}
\end{figure}

In Fig.~\ref{fig:comparison}(b) we give an example in which one bit of the start key is wrong. Although the image recovered by DGI with one bit of the start matrix incorrect (Fig.~\ref{fig:comparison} (b)) looks almost the same as that from the all-correct matrix in (a), the recovered keys (Figs.~\ref{fig:comparison} (d) and (c), respectively) are totally different. A miss is as good as a mile. This phenomenon is somewhat similar to a chaotic system which is sensitive to initial conditions; an arbitrarily small perturbation of the trajectory may lead to significantly different future behavior. Figure 5~(e) shows the CS image retrieved from the correct start key, while 5(f) is the CS image retrieved with 19 bits of the measured values incorrect, both from a total of 7500 frames. Here we have supposed each measured value to be one integer consisting of 4-bytes, with 8 bits in one byte, i.e. each integer contains 32 bits. In transmission, bit errors will occur with the same probability whether in high or low byte places, but errors in the former will have a more serious effect on the recovered image. In Fig.~5 (f) we have distributed the 19 bit errors evenly amongst the $7500 \times 32$ bits, and we can see that the image is completely indistinguishable. The bit error rate (BER) is about 0.08\%. Thus all parties should keep the BER under this tolerance level.

From the above analysis, we can regard this protocol as being perfectly safe for practical purposes.

2) Interception and falsification

In this method of attack Eve intercepts all measurement data, then forges a set of measurement data and sends it to all the other parties, thus she impersonates Alice and has the same status as the rightful recipients. But in the subsequent authentication and public comparison stage, the legitimate parties can easily discover her presence. If a party finds the CS recovered image to be different from the object provided beforehand, he/she can conclude that there is an eavesdropper. Here authentication via CS improves the security of the protocol. It is a necessary condition but not sufficient for guaranteeing the security of the protocol. After that, they can publicly compare a random subset of the generated key bits on which they should all agree. If they find few or no errors, the remaining uncompared data is secure enough to be a one-time-pad key.

\section{Conclusion}
In summary, we have demonstrated a one-time-pad key distribution protocol based on computational GI and compressed sensing for a point-to-multipoint optical network. It is only necessary to predistribute a secure matrix amongst the legitimate parties, from which they can then generate a key of completely random bits over several rounds of communication. The security of this privacy amplification and generation protocol is better than any classical cryptographic key distribution scheme, for it cannot be learnt or predicted. Some essential secure parameters are added to improve security, such as the number of measurements, and the imaging area. The start key can be distributed first between two cities through a QKD channel, then amplified and shared with other users in each city network. Compared with our former GI protocol, CS imaging provides faster authentication, increasing the key generation rate by an order of magnitude. With these advantages, this protocol may be widely used in inter-city secure communication and efficient conference key distribution systems, as well as electronic payment systems and other high security applications.

\bigskip
This work was supported by the National High Technology Research and Development Program of China (Grant No. 2011AA120102), the State Key Development Program for Basic Research of China (Grant No. 2010CB922904), and the National Natural Science Foundation of China (Grant Nos. 61178010 and 61274024).


\begin{thebibliography}{99}

\makeatletter
\providecommand \@ifxundefined [1]{%
 \@ifx{#1\undefined}
}%
\providecommand \@ifnum [1]{%
 \ifnum #1\expandafter \@firstoftwo
 \else \expandafter \@secondoftwo
 \fi
}%
\providecommand \@ifx [1]{%
 \ifx #1\expandafter \@firstoftwo
 \else \expandafter \@secondoftwo
 \fi
}%
\providecommand \natexlab [1]{#1}%
\providecommand \enquote  [1]{``#1''}%
\providecommand \bibnamefont  [1]{#1}%
\providecommand \bibfnamefont [1]{#1}%
\providecommand \citenamefont [1]{#1}%
\providecommand \href@noop [0]{\@secondoftwo}%
\providecommand \href [0]{\begingroup \@sanitize@url \@href}%
\providecommand \@href[1]{\@@startlink{#1}\@@href}%
\providecommand \@@href[1]{\endgroup#1\@@endlink}%
\providecommand \@sanitize@url [0]{\catcode `\\12\catcode `\$12\catcode
  `\&12\catcode `\#12\catcode `\^12\catcode `\_12\catcode `\%12\relax}%
\providecommand \@@startlink[1]{}%
\providecommand \@@endlink[0]{}%
\providecommand \url  [0]{\begingroup\@sanitize@url \@url }%
\providecommand \@url [1]{\endgroup\@href {#1}{\urlprefix }}%
\providecommand \urlprefix  [0]{URL }%
\providecommand \Eprint [0]{\href }%
\providecommand \doibase [0]{http://dx.doi.org/}%
\providecommand \selectlanguage [0]{\@gobble}%
\providecommand \bibinfo  [0]{\@secondoftwo}%
\providecommand \bibfield  [0]{\@secondoftwo}%
\providecommand \translation [1]{[#1]}%
\providecommand \BibitemOpen [0]{}%
\providecommand \bibitemStop [0]{}%
\providecommand \bibitemNoStop [0]{.\EOS\space}%
\providecommand \EOS [0]{\spacefactor3000\relax}%
\providecommand \BibitemShut  [1]{\csname bibitem#1\endcsname}%
\let\auto@bib@innerbib\@empty
\bibitem [{\citenamefont {Bennett}\ and\ \citenamefont {Brassard}(1984)}]{Bennett1984}%
  \BibitemOpen
  \bibfield  {author} {\bibinfo {author} {\bibfnamefont {C.~H.}\ \bibnamefont
  {Bennett}}\ and\ \bibinfo {author} {\bibfnamefont {G.}~\bibnamefont
  {Brassard}},\ }\bibfield  {title} {\enquote {\bibinfo {title} {Quantum
  cryptography: public-key distribution and coin tossing},}\ }in\ \href@noop {}
  {\emph {\bibinfo {booktitle} {Proc. ICCSSP}}}\ (\bibinfo {address}
  {Bangalore, India},\ \bibinfo {year} {1984})\ p.\ \bibinfo {pages}
  {175}\BibitemShut {NoStop}%
\bibitem [{\citenamefont {Bennett}\ \emph {et~al.}(1992)\citenamefont
  {Bennett}, \citenamefont {Bessette}, \citenamefont {Brassard}, \citenamefont
  {Salvail},\ and\ \citenamefont {Smolin}}]{Bennett1992}%
  \BibitemOpen
  \bibfield  {author} {\bibinfo {author} {\bibfnamefont {C.~H.}\ \bibnamefont
  {Bennett}}, \bibinfo {author} {\bibfnamefont {F.}~\bibnamefont {Bessette}},
  \bibinfo {author} {\bibfnamefont {G.}~\bibnamefont {Brassard}}, \bibinfo
  {author} {\bibfnamefont {L.}~\bibnamefont {Salvail}}, \ and\ \bibinfo
  {author} {\bibfnamefont {J.}~\bibnamefont {Smolin}},\ }\bibfield  {title}
  {\enquote {\bibinfo {title} {Experimental quantum cryptography},}\
  }\href@noop {} {\bibfield  {journal} {\bibinfo  {journal} {Journal of Cryp.}\
  }\textbf {\bibinfo {volume} {5}},\ \bibinfo {pages} {3} (\bibinfo {year}
  {1992})}\BibitemShut {NoStop}%
\bibitem [{\citenamefont {Bennett}, \citenamefont {Brassard},\ and\
  \citenamefont {Mermin}(1992)}]{Brassard1992}%
  \BibitemOpen
  \bibfield  {author} {\bibinfo {author} {\bibfnamefont {C.~H.}\ \bibnamefont
  {Bennett}}, \bibinfo {author} {\bibfnamefont {G.}~\bibnamefont {Brassard}}, \
  and\ \bibinfo {author} {\bibfnamefont {N.~D.}\ \bibnamefont {Mermin}},\
  }\bibfield  {title} {\enquote {\bibinfo {title} {Quantum cryptography without
  bell's theorem},}\ }\href@noop {} {\bibfield  {journal} {\bibinfo  {journal}
  {Phys. Rev. Lett.}\ }\textbf {\bibinfo {volume} {68}},\ \bibinfo {pages}
  {557} (\bibinfo {year} {1992})}\BibitemShut {NoStop}%
\bibitem [{\citenamefont {Bennett}, \citenamefont {Brassard},\ and\
  \citenamefont {Ekert}(1992)}]{Ekert1992}%
  \BibitemOpen
  \bibfield  {author} {\bibinfo {author} {\bibfnamefont {C.~H.}\ \bibnamefont
  {Bennett}}, \bibinfo {author} {\bibfnamefont {G.}~\bibnamefont {Brassard}}, \
  and\ \bibinfo {author} {\bibfnamefont {A.}~\bibnamefont {Ekert}},\ }\bibfield
   {title} {\enquote {\bibinfo {title} {Quantum cryptography},}\ }\href@noop {}
  {\bibfield  {journal} {\bibinfo  {journal} {Scientific Am.}\ }\textbf
  {\bibinfo {volume} {267}},\ \bibinfo {pages} {26} (\bibinfo {year}
  {1992})}\BibitemShut {NoStop}%
\bibitem [{\citenamefont {Lo}, \citenamefont {Chau},\ and\ \citenamefont
  {Ardehali}(2005)}]{Lo2005}%
  \BibitemOpen
  \bibfield  {author} {\bibinfo {author} {\bibfnamefont {H.~K.}\ \bibnamefont
  {Lo}}, \bibinfo {author} {\bibfnamefont {H.~F.}\ \bibnamefont {Chau}}, \ and\
  \bibinfo {author} {\bibfnamefont {M.}~\bibnamefont {Ardehali}},\ }\bibfield
  {title} {\enquote {\bibinfo {title} {Efficient quantum key distribution
  scheme and a proof of its unconditional security},}\ }\href@noop {}
  {\bibfield  {journal} {\bibinfo  {journal} {J. Cryptol.}\ }\textbf {\bibinfo
  {volume} {18 (2)}},\ \bibinfo {pages} {133} (\bibinfo {year}
  {2005})}\BibitemShut {NoStop}%
\bibitem [{\citenamefont {Hwang}, \citenamefont {Koh},\ and\ \citenamefont
  {Han}(1998)}]{Hwang1998}%
  \BibitemOpen
  \bibfield  {author} {\bibinfo {author} {\bibfnamefont {W.~Y.}\ \bibnamefont
  {Hwang}}, \bibinfo {author} {\bibfnamefont {I.~G.}\ \bibnamefont {Koh}}, \
  and\ \bibinfo {author} {\bibfnamefont {Y.~D.}\ \bibnamefont {Han}},\
  }\bibfield  {title} {\enquote {\bibinfo {title} {Quantum cryptography without
  public announcement of bases},}\ }\href@noop {} {\bibfield  {journal}
  {\bibinfo  {journal} {Phys. Rev. A}\ }\textbf {\bibinfo {volume} {244 (6)}},\
  \bibinfo {pages} {489} (\bibinfo {year} {1998})}\BibitemShut {NoStop}%
\bibitem [{\citenamefont {Guan}, \citenamefont {Wang},\ and\ \citenamefont
  {Zhang}(2012)}]{Guan2012}%
  \BibitemOpen
  \bibfield  {author} {\bibinfo {author} {\bibfnamefont {D.~J.}\ \bibnamefont
  {Guan}}, \bibinfo {author} {\bibfnamefont {Y.~J.}\ \bibnamefont {Wang}}, \
  and\ \bibinfo {author} {\bibfnamefont {E.~S.}\ \bibnamefont {Zhang}},\
  }\bibfield  {title} {\enquote {\bibinfo {title} {Quantum key evolution and
  its applications},}\ }\href@noop {} {\bibfield  {journal} {\bibinfo
  {journal} {Int. J. Quantum Inform.}\ }\textbf {\bibinfo {volume} {10}},\
  \bibinfo {pages} {1250044} (\bibinfo {year} {2012})}\BibitemShut {NoStop}%
\bibitem [{\citenamefont {Pittman}\ \emph {et~al.}(1995)\citenamefont
  {Pittman}, \citenamefont {Shih}, \citenamefont {Strekalov},\ and\
  \citenamefont {Sergienko}}]{Pittman1995}%
  \BibitemOpen
  \bibfield  {author} {\bibinfo {author} {\bibfnamefont {T.~B.}\ \bibnamefont
  {Pittman}}, \bibinfo {author} {\bibfnamefont {Y.~H.}\ \bibnamefont {Shih}},
  \bibinfo {author} {\bibfnamefont {D.~V.}\ \bibnamefont {Strekalov}}, \ and\
  \bibinfo {author} {\bibfnamefont {A.~V.}\ \bibnamefont {Sergienko}},\
  }\bibfield  {title} {\enquote {\bibinfo {title} {Optical imaging by means of
  two-photon quantum entanglement},}\ }\href@noop {} {\bibfield  {journal}
  {\bibinfo  {journal} {Phys. Rev. A}\ }\textbf {\bibinfo {volume} {52}},\
  \bibinfo {pages} {R3429} (\bibinfo {year} {1995})}\BibitemShut {NoStop}%
\bibitem [{\citenamefont {Gatti}\ \emph {et~al.}(2004)\citenamefont {Gatti},
  \citenamefont {Brambilla}, \citenamefont {Bache},\ and\ \citenamefont
  {Lugiato}}]{Lugiato2004}%
  \BibitemOpen
  \bibfield  {author} {\bibinfo {author} {\bibfnamefont {A.}~\bibnamefont
  {Gatti}}, \bibinfo {author} {\bibfnamefont {E.}~\bibnamefont {Brambilla}},
  \bibinfo {author} {\bibfnamefont {M.}~\bibnamefont {Bache}}, \ and\ \bibinfo
  {author} {\bibfnamefont {L.~A.}\ \bibnamefont {Lugiato}},\ }\bibfield
  {title} {\enquote {\bibinfo {title} {Ghost imaging with thermal light:
  comparing entanglement and classical correlation},}\ }\href@noop {}
  {\bibfield  {journal} {\bibinfo  {journal} {Phys. Rev. Lett.}\ }\textbf
  {\bibinfo {volume} {93}},\ \bibinfo {pages} {093602} (\bibinfo {year}
  {2004})}\BibitemShut {NoStop}%
\bibitem [{\citenamefont {Zhang}\ \emph {et~al.}(2005)\citenamefont {Zhang},
  \citenamefont {Zhai}, \citenamefont {Wu},\ and\ \citenamefont
  {Chen}}]{LAWu2005}%
  \BibitemOpen
  \bibfield  {author} {\bibinfo {author} {\bibfnamefont {D.}~\bibnamefont
  {Zhang}}, \bibinfo {author} {\bibfnamefont {Y.~H.}\ \bibnamefont {Zhai}},
  \bibinfo {author} {\bibfnamefont {L.~A.}\ \bibnamefont {Wu}}, \ and\ \bibinfo
  {author} {\bibfnamefont {X.~H.}\ \bibnamefont {Chen}},\ }\bibfield  {title}
  {\enquote {\bibinfo {title} {Correlated two-photon imaging with true thermal
  light},}\ }\href@noop {} {\bibfield  {journal} {\bibinfo  {journal} {Opt.
  Lett.}\ }\textbf {\bibinfo {volume} {30}},\ \bibinfo {pages} {2354} (\bibinfo
  {year} {2005})}\BibitemShut {NoStop}%
\bibitem [{\citenamefont {Erkmen}\ and\ \citenamefont
  {Shapiro}(2008)}]{Erkmen2008}%
  \BibitemOpen
  \bibfield  {author} {\bibinfo {author} {\bibfnamefont {B.~I.}\ \bibnamefont
  {Erkmen}}\ and\ \bibinfo {author} {\bibfnamefont {J.~H.}\ \bibnamefont
  {Shapiro}},\ }\bibfield  {title} {\enquote {\bibinfo {title} {Unified theory
  of ghost imaging with {Gaussian-state} light},}\ }\href@noop {} {\bibfield
  {journal} {\bibinfo  {journal} {Phys. Rev. A}\ }\textbf {\bibinfo {volume}
  {77}},\ \bibinfo {pages} {043809} (\bibinfo {year} {2008})}\BibitemShut
  {NoStop}%
\bibitem [{\citenamefont {Brida}\ \emph {et~al.}(2011)\citenamefont {Brida},
  \citenamefont {Chekhova}, \citenamefont {Fornaro}, \citenamefont {Genovese},
  \citenamefont {Lopaeva},\ and\ \citenamefont {Berchera}}]{Brida2011}%
  \BibitemOpen
  \bibfield  {author} {\bibinfo {author} {\bibfnamefont {G.}~\bibnamefont
  {Brida}}, \bibinfo {author} {\bibfnamefont {M.~V.}\ \bibnamefont {Chekhova}},
  \bibinfo {author} {\bibfnamefont {G.~A.}\ \bibnamefont {Fornaro}}, \bibinfo
  {author} {\bibfnamefont {M.}~\bibnamefont {Genovese}}, \bibinfo {author}
  {\bibfnamefont {E.~D.}\ \bibnamefont {Lopaeva}}, \ and\ \bibinfo {author}
  {\bibfnamefont {I.~R.}\ \bibnamefont {Berchera}},\ }\bibfield  {title}
  {\enquote {\bibinfo {title} {Systematic analysis of signal-to-noise ratio in
  bipartite ghost imaging with classical and quantum light},}\ }\href@noop {}
  {\bibfield  {journal} {\bibinfo  {journal} {Phys. Rev. A}\ }\textbf {\bibinfo
  {volume} {83}},\ \bibinfo {pages} {063807} (\bibinfo {year}
  {2011})}\BibitemShut {NoStop}%
\bibitem [{\citenamefont {Shapiro}(2008)}]{Shapiro2008}%
  \BibitemOpen
  \bibfield  {author} {\bibinfo {author} {\bibfnamefont {J.~H.}\ \bibnamefont
  {Shapiro}},\ }\bibfield  {title} {\enquote {\bibinfo {title} {Computational
  ghost imaging},}\ }\href@noop {} {\bibfield  {journal} {\bibinfo  {journal}
  {Phy. Rev. A}\ }\textbf {\bibinfo {volume} {78}},\ \bibinfo {pages} {061802R}
  (\bibinfo {year} {2008})}\BibitemShut {NoStop}%
\bibitem [{\citenamefont {Clemente}\ \emph {et~al.}(2010)\citenamefont
  {Clemente}, \citenamefont {Dur$\acute{a}$n}, \citenamefont {Torres-Company},
  \citenamefont {Tajahuerce},\ and\ \citenamefont {Lanci}}]{Clemente2010}%
  \BibitemOpen
  \bibfield  {author} {\bibinfo {author} {\bibfnamefont {P.}~\bibnamefont
  {Clemente}}, \bibinfo {author} {\bibfnamefont {V.}~\bibnamefont
  {Dur$\acute{a}$n}}, \bibinfo {author} {\bibfnamefont {V.}~\bibnamefont
  {Torres-Company}}, \bibinfo {author} {\bibfnamefont {E.}~\bibnamefont
  {Tajahuerce}}, \ and\ \bibinfo {author} {\bibfnamefont {J.}~\bibnamefont
  {Lanci}},\ }\bibfield  {title} {\enquote {\bibinfo {title} {Optical
  encryption based on computational ghost imaging},}\ }\href@noop {} {\bibfield
   {journal} {\bibinfo  {journal} {Opt. Lett.}\ }\textbf {\bibinfo {volume}
  {35}},\ \bibinfo {pages} {2391} (\bibinfo {year} {2010})}\BibitemShut
  {NoStop}%
\bibitem [{\citenamefont {Tanha}, \citenamefont {Kheradmand},\ and\
  \citenamefont {Ahmadi-Kandjan}(2012)}]{Tanha2012}%
  \BibitemOpen
  \bibfield  {author} {\bibinfo {author} {\bibfnamefont {M.}~\bibnamefont
  {Tanha}}, \bibinfo {author} {\bibfnamefont {R.}~\bibnamefont {Kheradmand}}, \
  and\ \bibinfo {author} {\bibfnamefont {S.}~\bibnamefont {Ahmadi-Kandjan}},\
  }\bibfield  {title} {\enquote {\bibinfo {title} {Gray-scale and color optical
  encryption based on computational ghost imaging},}\ }\href@noop {} {\bibfield
   {journal} {\bibinfo  {journal} {Appl. Phys. Lett.}\ }\textbf {\bibinfo
  {volume} {101}},\ \bibinfo {pages} {101108} (\bibinfo {year}
  {2012})}\BibitemShut {NoStop}%
\bibitem [{\citenamefont {Donoho}(2006)}]{Donoho2006}%
  \BibitemOpen
  \bibfield  {author} {\bibinfo {author} {\bibfnamefont {D.}~\bibnamefont
  {Donoho}},\ }\bibfield  {title} {\enquote {\bibinfo {title} {Compressed
  sensing},}\ }\href@noop {} {\bibfield  {journal} {\bibinfo  {journal} {IEEE
  Trans. Inform. Theory}\ }\textbf {\bibinfo {volume} {52 (4)}},\ \bibinfo
  {pages} {1289} (\bibinfo {year} {2006})}\BibitemShut {NoStop}%
\bibitem [{\citenamefont {Cand$\grave{e}$s}(2006)}]{Candes2006}%
  \BibitemOpen
  \bibfield  {author} {\bibinfo {author} {\bibfnamefont {E.~J.}\ \bibnamefont
  {Cand$\grave{e}$s}},\ }\bibfield  {title} {\enquote {\bibinfo {title}
  {Compressive sampling},}\ }in\ \href@noop {} {\emph {\bibinfo {booktitle}
  {Proc. Int. Cong. Math., Spain}}},\ \bibinfo {series and number} {3}\
  (\bibinfo {organization} {European Mathematical Society},\ \bibinfo {address}
  {Madrid, Spain},\ \bibinfo {year} {2006})\ p.\ \bibinfo {pages}
  {1433}\BibitemShut {NoStop}%
\bibitem [{\citenamefont {Duarte}\ \emph {et~al.}(2008)\citenamefont {Duarte},
  \citenamefont {Davenport}, \citenamefont {Takhar}, \citenamefont {Laska},
  \citenamefont {Sun}, \citenamefont {Kelly},\ and\ \citenamefont
  {Baraniuk}}]{Baraniuk2008}%
  \BibitemOpen
  \bibfield  {author} {\bibinfo {author} {\bibfnamefont {M.~F.}\ \bibnamefont
  {Duarte}}, \bibinfo {author} {\bibfnamefont {M.~A.}\ \bibnamefont
  {Davenport}}, \bibinfo {author} {\bibfnamefont {D.}~\bibnamefont {Takhar}},
  \bibinfo {author} {\bibfnamefont {J.~N.}\ \bibnamefont {Laska}}, \bibinfo
  {author} {\bibfnamefont {T.}~\bibnamefont {Sun}}, \bibinfo {author}
  {\bibfnamefont {K.~F.}\ \bibnamefont {Kelly}}, \ and\ \bibinfo {author}
  {\bibfnamefont {R.~G.}\ \bibnamefont {Baraniuk}},\ }\bibfield  {title}
  {\enquote {\bibinfo {title} {Single-pixel imaging via compressive
  sampling},}\ }\href@noop {} {\bibfield  {journal} {\bibinfo  {journal} {IEEE
  Sig. Proc. Magazine}\ }\textbf {\bibinfo {volume} {25 (2)}},\ \bibinfo
  {pages} {83} (\bibinfo {year} {2008})}\BibitemShut {NoStop}%
\bibitem [{\citenamefont {Chan}\ \emph {et~al.}(2008)\citenamefont {Chan},
  \citenamefont {Charan}, \citenamefont {Takhar}, \citenamefont {Kelly},
  \citenamefont {Baraniuk},\ and\ \citenamefont {Mittleman}}]{Daniel2008}%
  \BibitemOpen
  \bibfield  {author} {\bibinfo {author} {\bibfnamefont {W.~L.}\ \bibnamefont
  {Chan}}, \bibinfo {author} {\bibfnamefont {K.}~\bibnamefont {Charan}},
  \bibinfo {author} {\bibfnamefont {D.}~\bibnamefont {Takhar}}, \bibinfo
  {author} {\bibfnamefont {K.~F.}\ \bibnamefont {Kelly}}, \bibinfo {author}
  {\bibfnamefont {R.~G.}\ \bibnamefont {Baraniuk}}, \ and\ \bibinfo {author}
  {\bibfnamefont {D.~M.}\ \bibnamefont {Mittleman}},\ }\bibfield  {title}
  {\enquote {\bibinfo {title} {A single-pixel terahertz imaging system based on
  compressed sensing},}\ }\href@noop {} {\bibfield  {journal} {\bibinfo
  {journal} {Appl. Phys. Lett.}\ }\textbf {\bibinfo {volume} {93}},\ \bibinfo
  {pages} {121105} (\bibinfo {year} {2008})}\BibitemShut {NoStop}%
\bibitem [{\citenamefont {Katz}, \citenamefont {Bromberg},\ and\ \citenamefont
  {Silberberg}(2009)}]{Silberberg2009}%
  \BibitemOpen
  \bibfield  {author} {\bibinfo {author} {\bibfnamefont {O.}~\bibnamefont
  {Katz}}, \bibinfo {author} {\bibfnamefont {Y.}~\bibnamefont {Bromberg}}, \
  and\ \bibinfo {author} {\bibfnamefont {Y.}~\bibnamefont {Silberberg}},\
  }\bibfield  {title} {\enquote {\bibinfo {title} {Compressive ghost
  imaging},}\ }\href@noop {} {\bibfield  {journal} {\bibinfo  {journal} {Appl.
  Phys. Lett.}\ }\textbf {\bibinfo {volume} {95}},\ \bibinfo {pages} {131110}
  (\bibinfo {year} {2009})}\BibitemShut {NoStop}%
\bibitem [{\citenamefont {Li}\ \emph {et~al.}(2013)\citenamefont {Li},
  \citenamefont {Yao}, \citenamefont {Yu}, \citenamefont {Wu},\ and\
  \citenamefont {Zhai}}]{LiShen2013}%
  \BibitemOpen
  \bibfield  {author} {\bibinfo {author} {\bibfnamefont {S.}~\bibnamefont
  {Li}}, \bibinfo {author} {\bibfnamefont {X.~R.}\ \bibnamefont {Yao}},
  \bibinfo {author} {\bibfnamefont {W.~K.}\ \bibnamefont {Yu}}, \bibinfo
  {author} {\bibfnamefont {L.~A.}\ \bibnamefont {Wu}}, \ and\ \bibinfo {author}
  {\bibfnamefont {G.~J.}\ \bibnamefont {Zhai}},\ }\bibfield  {title} {\enquote
  {\bibinfo {title} {High-speed secure key distribution over an optical network
  based on computational correlation imaging},}\ }\href@noop {} {\bibfield
  {journal} {\bibinfo  {journal} {Opt. Lett.}\ }\textbf {\bibinfo {volume} {38
  (12)}},\ \bibinfo {pages} {2144} (\bibinfo {year} {2013})}\BibitemShut
  {NoStop}%
\bibitem [{\citenamefont {Ferri}\ \emph {et~al.}(2010)\citenamefont {Ferri},
  \citenamefont {Magatti}, \citenamefont {Lugiato},\ and\ \citenamefont
  {Gatti}}]{Gatti2010}%
  \BibitemOpen
  \bibfield  {author} {\bibinfo {author} {\bibfnamefont {F.}~\bibnamefont
  {Ferri}}, \bibinfo {author} {\bibfnamefont {D.}~\bibnamefont {Magatti}},
  \bibinfo {author} {\bibfnamefont {L.~A.}\ \bibnamefont {Lugiato}}, \ and\
  \bibinfo {author} {\bibfnamefont {A.}~\bibnamefont {Gatti}},\ }\bibfield
  {title} {\enquote {\bibinfo {title} {Differential ghost imaging},}\
  }\href@noop {} {\bibfield  {journal} {\bibinfo  {journal} {Phys. Rev. Lett.}\
  }\textbf {\bibinfo {volume} {104}},\ \bibinfo {pages} {253603} (\bibinfo
  {year} {2010})}\BibitemShut {NoStop}%
\bibitem [{\citenamefont {Li}(2010)}]{Li2010}%
  \BibitemOpen
  \bibfield  {author} {\bibinfo {author} {\bibfnamefont {C.~B.}\ \bibnamefont
  {Li}},\ }\emph {\bibinfo {title} {An efficient algorithm for total variation
  regularization with applications to the single pixel camera and compressive
  sensing}},\ \href@noop {} {\bibinfo {type} {M.{S}. thesis}},\ \bibinfo
  {school} {Rice University} (\bibinfo {year} {2010})\BibitemShut {NoStop}%
\bibitem [{\citenamefont {Walker}(2008)}]{Walker2008}%
  \BibitemOpen
  \bibfield  {author} {\bibinfo {author} {\bibfnamefont {J.}~\bibnamefont
  {Walker}},\ }\href@noop {} {\enquote {\bibinfo {title} {Ent: a pseudorandom
  number sequence test program},}\ }\bibinfo {type} {Tech. Rep.}\ (\bibinfo
  {address} \url{http://www.fourmilab.ch/random},\ \bibinfo {year}
  {2008})\BibitemShut {NoStop}%
\bibitem [{\citenamefont {Instruments}(2013{\natexlab{a}})}]{TI1080}%
  \BibitemOpen
  \bibfield  {author} {\bibinfo {author} {\bibfnamefont {Texas Instruments}},\ }\href@noop {} {\enquote {\bibinfo {title} {{DLP} 0.95 1080p
  type {A} {DMD} ({Rev. B})},}\ }\bibinfo {type} {Tech. Rep.}\ (\bibinfo
  {address} {http://www.ti.com/lit/ds/dlps025b/dlps025b.pdf},\ \bibinfo {year}
  {2013})\BibitemShut {NoStop}%
\bibitem [{\citenamefont {Instruments}(2013{\natexlab{b}})}]{TI4100}%
  \BibitemOpen
  \bibfield  {author} {\bibinfo {author} {\bibfnamefont {Texas Instruments}},\ }\href@noop {} {\enquote {\bibinfo {title} {{DLP} discovery
  4100 chipset data sheet ({Rev. A})},}\ }\bibinfo {type} {Tech. Rep.}\
  (\bibinfo {address} {http://www.ti.com/lit/er/dlpu008a/dlpu008a.pdf},\
  \bibinfo {year} {2013})\BibitemShut {NoStop}%
\end{thebibliography}
\end{document}